\documentclass{agujournal2019}
\usepackage{url} 
\usepackage[inline]{trackchanges} 
\usepackage{soul}
\usepackage{graphicx} 
\usepackage{comment}
\usepackage{amsmath}
\usepackage{soul}
\usepackage{appendix}
\usepackage[round]{natbib}
\usepackage{xcolor}
\setcitestyle{authoryear,open={(},close={)}}
\newcommand{\beq}{\begin{eqnarray}}
\newcommand{\eeq}{\end{eqnarray}}
\linenumbers
\draftfalse

\journalname{Geophysical Research Letters}

\nolinenumbers 

\begin{document}

%
%


\title{Analytical Model for the Higher Order  Moments of Midlatitude Atmospheric Temperature Distributions}

%
%




\authors{Keiko Kircher, Cristi Proistosescu, and Ryan L. Sriver}


\affiliation{}{Department of Climate, Meteorology, and Atmospheric Sciences, University of Illinois at Urbana-Champaign}




\correspondingauthor{Keiko Kircher}{keikoino@illinois.edu}




\begin{keypoints}
\item An analytical expression of the higher order moments of atmospheric temperature distributions is derived 
\item Latitudinal nonlinearity in temperature explains skew and kurtosis in mid-latitudes for Held-Suarez model with current temperature profile
\item Changes in wind and temperature profiles equally contribute to decreases in moments due to climate change


\end{keypoints}

%
%

%
%


\begin{abstract}

Observed distributions of atmospheric temperature are non-Gaussian. Therefore, moments beyond variance are necessary in determining the frequency of extreme temperature events. Here we propose a simple kinematic model for atmospheric mid-latitude temperature variability based on symmetric advection from a non-symmetric background temperature profile. We then use this model to derive analytical expressions for the higher order moments of  temperature distributions. Our results show that nonzero skewness and kurtosis arise due to the nonlinearity of the time-mean meridional temperature profile. The analytical model matches an idealized Held-suarez atmospheric model, indicating nonlinearity of time-mean temperature in latitude is the dominant contribution to nonzero skewness and kurtosis in synoptic temperature variations. Model analysis further shows decrease in higher order moments due to climate change come roughly equally from changes in mixing length and changes in the background temperature profiles.



\end{abstract}

\section*{Plain Language Summary}

As climate change increases the average temperature of the Earth, the frequency and intensity of extreme temperature events changes as well. Understanding these changing extremes requires more than just understanding how the mean temperature increases; it requires an understanding of the entire distribution of daily temperatures and how this distribution changes. In this work, we build a theory for how  the distribution of daily temperature is related to the time-average temperature, and test this theory in a numerical climate model. We show that temperature changes \textit{in time} at one fixed location are related to how the time-mean temperature varies \textit{in space}. Intuitively, this means we experience hot days when the air above us comes from a region where it is hotter \textit{on average}, and we experience cold days when the air above us comes from a region where it is colder \textit{on average}. Thus, we link a quantity that is important but difficult to measure - the frequency of temperature extremes -  to a quantity that is well measured and more easily projectable into the future - the time-average (or climatological) temperature profile.


%
%

%


%
%
%
%

\section{Introduction}


Observations show that atmospheric distribution of temperature is non-Gaussian \citep{white1980skewness,trenberth1985blocking,nakamura1991skewness,holzer1996asymmetric}. This non-Gaussianity implies that mean and variance of temperature variability are not sufficient in determining the distribution of temperature extremes, and consideration of higher-order moments is essential. Therefore, understanding what controls higher-order moments in temperature distribution is an important part of understanding the frequency of extreme temperature anomalies, such as heat waves and cold spells, and how this frequency might change in a changing climate. 


Here we focus on synoptic sources of temperature variability in midlatitudes, while understanding that temperatures in the surface boundary layer are also impacted by other processes. Such processes include convective coupling of the surface to the free troposphere \citep[e.g.,][]{zhang2023upper,byrne2021amplified}, diabatic heating associated with vertical motion during heat-dome events \citep[e.g.,][]{white2023unprecedented} and the effects of coupling to the land surface and soil moisture  \citep[e.g.,][]{seneviratne2010investigating,zeppetello2022physics}.

To model synoptic temperature variability in the mid-latitude lower troposphere, a standard framework assumes that the temperature is controlled by Lagrangian advection from a background temperature gradient  \citep{schneider2015physics,garfinkel2017non,linz2018large,linz2020framework, tamarin2019dynamical,tamarin2020changes,tamarin2022simple,zhang2022interpreting}. This framework provides a straightforward prediction for the standard deviation $\sigma_T$ of lower-tropospheric (potential) temperature $T$; $\sigma_T$ is proportional to the mean meridional angular displacement of air $\eta$ and the first derivative of the time-mean (potential) temperature $\bar T$ in the meridional direction $\phi$ \citep{schneider2015physics}:  
\begin{equation}
\sigma_T ^2 = \left( \frac{d\bar T}{d\phi} \right)^2 \eta ^2, \label{eq:schneider}
\end{equation} 
where the average meridional displacement $\eta$ is assumed to be proportional to the Lagrangian mixing length \citep{corrsin1975limitations}.


A particular appeal of this framework is that it allows for partitioning \textit{changes} in temperature variance into contributions from changes in the time-mean temperature profile $\bar T(\phi)$ and changes in atmospheric circulation through their impacts on $\eta$:
\beq
\frac{\Delta \sigma_T ^2}{\sigma_T ^2} = \frac{\Delta (d\bar T/d\phi) ^2}{(d\bar T/d\phi) ^2} + \frac{\Delta \eta ^2}{\eta ^2}. \label{eq:deltasigma}
\eeq
Since the fractional change in the (square of) meridional mixing length $\Delta \eta^2/\eta ^2$ is expected to be much smaller than the fractional change in $(d \bar T / d\phi)^2$ under arctic amplification \citep{boer1983large, shepherd1987spectral, schneider2008scaling,merlis2009scales,o2011effective}, the decrease in $d \bar T / d\phi$ due to arctic amplification should be the dominant factor in the change in variance. This dominant factor and Eq. (\ref{eq:deltasigma}) lead to a prediction of a decrease in variance $\sigma_T^2$ due to climate change, which has been confirmed by comparisons of the theory with Reanalysis and climate model data in summer and winter at 850 hPa \citep{schneider2015physics} and at surface \citep{dai2021arctic}.


The framework has also been used for both theoretical \citep{luxford2012simple,tamarin2019dynamical,tamarin2020changes,tamarin2022simple, proistosescu2016identification} and numerical  \citep{garfinkel2017non,linz2018large,linz2020framework,zhang2023upper} explorations of higher order moments, mostly skewness. \citet{luxford2012simple} explained the observed pattern of negative skewness equatorward and positive skewness poleward of the mean jet stream \citep{white1980skewness,trenberth1985blocking}, using a kinematic model of jet movement. For an illustration of this idea, consider a location slightly equatorward from the average position of the jet stream. A poleward deviation of the jet stream further away from this location would result in a slight warm anomaly, whereas a deviation of the jet stream equatorward would result in much larger cold anomalies as the location finds itself poleward of the jet. Thus, symmetric deviations of the jet location would result in different intensities of cold and warm anomalies, which in turn would result in a skewed temperature distribution with skewness $S$ that is approximately given by
\begin{equation}
    S\approx \frac{T_w-T_c}{\sigma_T},
\end{equation}
where $T_w$, and $T_c$ denote the average absolute intensities of cold and warm anomalies, respectively \citep{tamarin2020changes}. By comparing this representation of skewness with historical CMIP data, \citet{tamarin2020changes} found that the primary control on the asymmetry of cold and warm anomalies comes from differences between the poleward and equatorward temperature gradient $\Delta (dT/d\phi)$, rather than differences between the poleward and equatorward mixing scales $\Delta \eta$, analogous to the case with variance as mentioned previously.

Here, we build on the Lagrangian advection framework of \citet{schneider2015physics} and \citet{tamarin2019dynamical,tamarin2020changes,tamarin2022simple} to provide theoretical predictions for the higher order moments of an arbitrary order that arise due to synoptic temperature variability. However, because the displacement length is limited by the distance to the pole in high latitudes, which can introduce additional skewness in temperature distributions, we focus on midlatitude variability in this work.
We propose a conceptual kinematic model where temperature at a given location is determined by a source temperature being advected from a nearby location. We use this model to calculate the probability distribution of the temperature and closed-form approximations of standard deviation, skewness, and kurtosis. We then show that our theoretical predictions for the moments are supported by an idealized dry-core atmospheric general circulation model (GCM) \citep[ISCA,][]{gmd-11-843-2018}. Finally, we show that our theory is able to predict and explain \textit{changes} in the higher order moments from climate change as either due to the change in the temperature profile or due to the change in mixing length.


\section{A conceptual model of stochastic advection}

\subsection{Kinematic representation}

\begin{figure}
\centering
\includegraphics[width=0.6\textwidth]{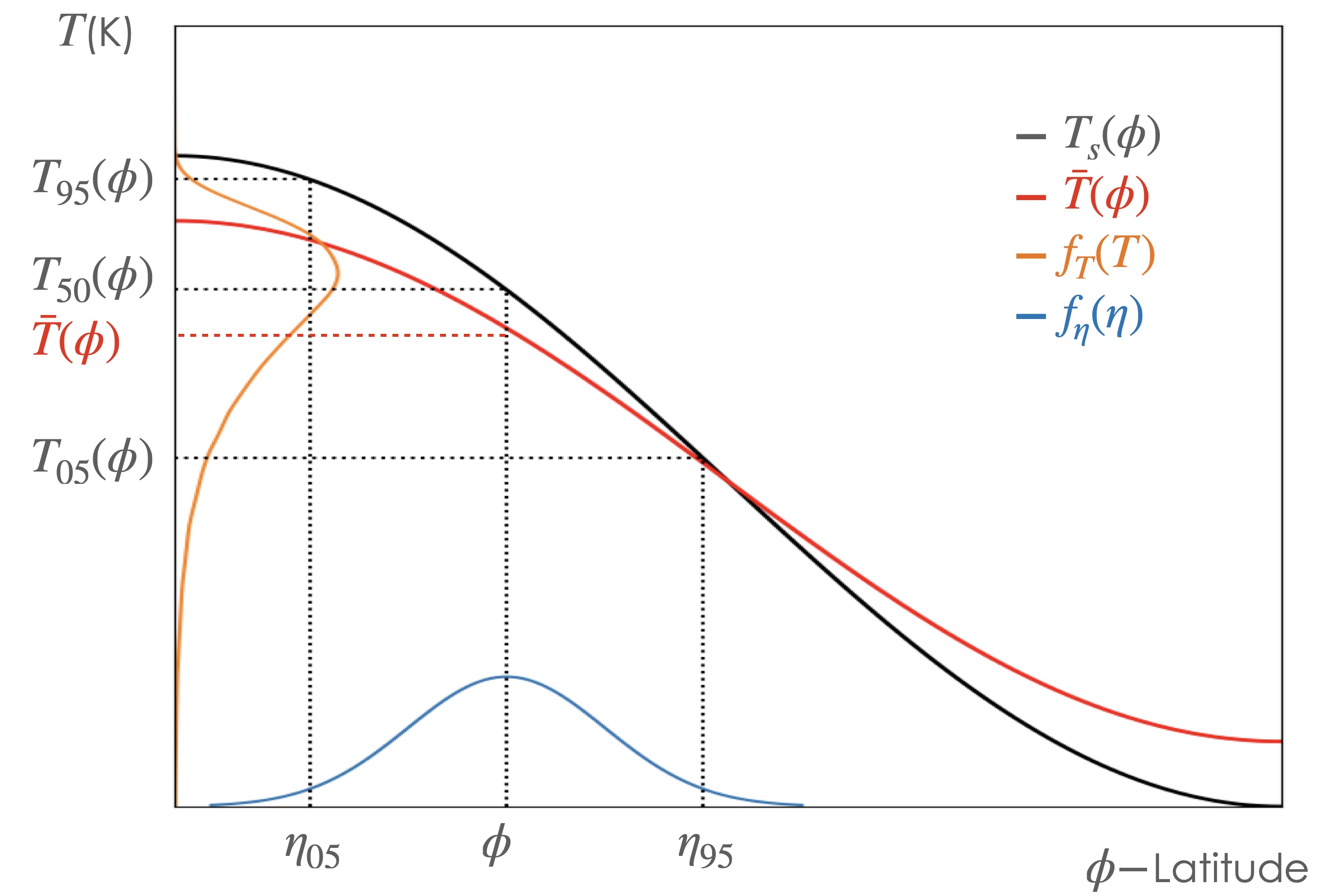}
    \caption{Graphical representations of the relationship between normal distribution of advective displacement and temperature distribution, $T(\phi,t)=T_s(\phi+\eta(t))$. When the equilibrium temperature is nonlinear in $\phi$, the resulting pdf in temperature is asymmetric even if the distribution of the displacement $\eta$ is symmetric.}
\label{fig:schematic}
\end{figure}

Our conceptual model is based on the assumption that atmospheric temperature at a given location is solely determined by the advection of a stationary source temperature via a stochastic wind with a Gaussian distribution. Specifically, we assume that the temperature at latitude $\phi$ and time $t$ is equal to the time-independent source temperature $T_s(\phi)$ advected from a nearby source latitude $\phi + \eta(t)$, i.e., 
\beq
T(\phi,t)=T_s(\phi+\eta(t)),\label{eq:kinematic}
\eeq
where the distribution of the displacement $\eta (t)$ is stochastic and normally distributed, 
\beq
\eta(t)\sim \mathcal N \left(0,\sigma^2_\eta \right)\label{eq:psi_dist},
\eeq
with probability density function $f_\eta(\eta)$. Since the time dependence comes solely from $\eta(t)$ as implied in Eq.(\ref{eq:kinematic}), the mean and median temperatures are stationary in this model. This model is similar to Schneider et al (2015), and can be considered the Brownian limit of stochastic advection in a Gaussian velocity field (O'Gormann and Schneider, 2006), i.e., the limit where the correlation timescale of the velocity field is much smaller than other relevant timescales, such as the radiative equilibration timescale. Similar models have been used in the past to explain the non-gaussianity of equilibrium climate sensitivity \citep{roe2007climate}. As shown schematically in Figure \ref{fig:schematic}, even though $f_\eta(\eta)$ is symmetric, if the temperature profile $T_s(\phi)$ is nonlinear in latitude (as is the case for the Earth), the resulting temperature distribution is asymmetric. The asymmetry of the temperature distribution in this setup is also clear from the rule for probability density functions $f_T(T)dT = f_\eta (\eta)d\eta$. This gives $f_T(T) = f_\eta(\eta)(d\eta /dT) = f_\eta(\eta)(d\phi /dT_s) = f_\eta(\eta) (dT_s/d\phi)^{-1}$ if $T$ is given by Eq. (\ref{eq:kinematic}). Clearly, if $T_s$ is nonlinear in $\phi$, $dT_s/d\phi$ is not a constant, which can make $f_T(T)$ asymmetric (unless $(dT_s/d\phi)^{-1}$ happens to be symmetric). Our objective in this work is to calculate the statistical moments of temperature that result from this asymmetry. Note that this argument is valid only if the derivative $d\phi/dT_s$ exists. Therefore, our theory is valid only away from locations where  $dT_s/d\phi = 0$, such as at the pole or the equator.

This kinematic representation illustrates a subtle deviation from the previous assumption that the advected temperature is the time-mean temperature \citep[e.g.,][]{schneider2015physics,tamarin2020changes}. As seen from the graphical representation in Fig. 1, the advected source temperature is the \emph{median} temperature. The first task then is to express the median temperature in terms of the mean and its derivatives, as shown in the next subsection.

\subsection{Calculation of moments} \label{Calculation of moments}

In this section, we use Eq. (\ref{eq:kinematic}) as a mapping between temperature $T$ and latitude $\phi$ to calculate the moments of temperature in terms of a known time-mean temperature profile $\bar T(\phi)$. To do so, we start with the definition of (normalized) moments,
\beq
\mu_{k}= \frac{1}{ \sigma_T^k} \int (T-\bar T)^k f_T(T)dT, \label{eq:moments_T}
\eeq
where $f_T(T)$ is the probability density function of temperature, and $\bar T$ is the time-mean temperature,
\beq
\bar T = \int Tf_T(T)dT. \label{eq:mean_T}
\eeq
Using Eq. (\ref{eq:kinematic}) in Eqs. (\ref{eq:moments_T},\ref{eq:mean_T}), and keeping track of the change of variables rule for probability density functions, $f_T(T)dT=f_{\eta} (\eta)d\eta$, the moments at $\phi$ become
\beq
\mu _{k}(\phi)= \frac{1}{ \sigma_T^ k} \int (T_{s}(\phi+ \eta)-\bar T(\phi))^k f_{\eta}(\eta) d\eta \label{moments}
\eeq
and
\beq
\bar T (\phi)= \int T_{s}(\phi+\eta)f_\eta(\eta)d\eta, \label{barT}
\eeq
where $f_\eta (\eta)$ is the probability density function of gaussian random variable with mean zero and variance $\sigma^2_\eta$.

Since $T_s$ is not a priori known, we wish to express $T_{s}$ in terms of time-mean temperature $\bar T$, which is both measurable in present climates and predictable in future climates. For that, we calculate $\bar T$ by expanding $T_{s}(\phi+\eta)$ in Eq. (\ref{barT}) about $\phi$ as
\beq
\bar T (\phi)= \int \left(T_{s}(\phi) + \frac{dT_{s}}{d\phi}\bigg|_{\eta=0} \eta + \frac{1}{2!} \frac{d^2 T_{s}}{d\phi ^2} \bigg|_{\eta = 0} \eta ^2 + \frac{1}{3!} \frac{d^3T_{s}}{d\phi ^3} \bigg| _{\eta = 0} \eta ^3 + ... \right) f_\eta(\eta)d\eta. \label{Tbar}
\eeq
For Earth's atmosphere, the mixing length is much smaller than the length scale of the change in temperature gradient \citep{Swanson:1997,Daoud:2003},
\beq
\sigma_\eta << \frac{\partial T_{s}/\partial \phi}{\partial ^2 T_{s}/\partial 
 \phi^2} \label{eq:regime}.
\eeq
Therefore, it is reasonable to calculate $\bar T$ only to the lowest nonzero order in $\sigma _\eta$, as
\beq
\bar T (\phi)\approx T_{s} (\phi)+ \frac{1}{2!} \frac{d^2T_{s}}{d\phi ^2}\bigg|_{\eta = 0} \sigma_\eta ^2. \label{T-T}
\eeq
Conversely, to the lowest nonzero order in $\sigma_\eta$, the source field $T_s(\phi)$ can be written as:
\beq
T_{s}(\phi) \approx \bar T (\phi) - \frac{1}{2!} \frac{d^2\bar T}{d\phi ^2}\bigg|_{\eta = 0}\sigma_\eta ^2. \label{T-T2}
\eeq

Now Eq. (\ref{moments}) can be rewritten by expanding $T_s(\phi + \eta)$ and also by substituting Eq. (\ref{Tbar}) and Eq. (\ref{T-T2}) as 
\beq
\mu _k (\phi) = \frac{1}{\sigma ^k _T}\int \left(\bar T(\phi+\eta) - \frac{1}{2!} \frac{d^2\bar T}{d\phi ^2}\bigg|_{\eta = 0}\sigma_\eta ^2 - \bar T (\phi)\right)^kf_\eta(\eta)d\eta \label{eq:before_expansion} \\
=\frac{1}{\sigma_{T}^{k}}\int\left(\frac{d\bar{T}}{d\phi}\bigg|_{\eta = 0}\eta+\frac{1}{2!}\frac{d^{2}\bar{T}}{d\phi^{2}}\bigg|_{\eta = 0}(\eta^{2}-\sigma_\eta^{2})+\frac{d^{3}\bar{T}}{d\phi^{3}}\bigg|_{\eta = 0}\left(\frac{1}{3!}\eta^{3}-\frac{1}{2}\eta\sigma_\eta^{2}\right)+...\right)^{k}f_{\eta}(\eta)d\eta 
. \label{moments_expanded}
\eeq
Eq. (\ref{eq:before_expansion}) and (\ref{moments_expanded}) are applicable for moments of any order $k$ within our physical assumption. However, since the first four moments are of particular interest in most cases, we calculate the expressions of variance, skewness, and kurtosis to the lowest nonzero order in $\sigma_\eta$ explicitly: 
\beq
\sigma _T^2 \ &=& \ \left( \frac{d\bar T}{d\phi}\right)^2 \sigma_\eta ^2\label{sigmaT_theory} \\
\mu _{3} &=&3\left(\frac{d^2 \bar T}{d\phi ^2}\right) \frac{\sigma_\eta^2}{\sigma _T} \label{muT_theory} \\
\mu _{4} &=&3+\frac{5}{3}\mu_{3}^{2}+4\left(\frac{d\bar{T}}{d\phi}\right)\left(\frac{d^{3}\bar{T}}{d\phi^{3}}\right)\frac{\sigma_\eta^{4}}{\sigma_{T}^{2}}. \label{kurtosis}
\eeq
Just as in \cite{schneider2015physics}, standard deviation $\sigma _T$ decreases as pole-to-equator temperature decreases since it reduces the derivative $d\bar T/d\phi$. Skewness and excess kurtosis (i.e., $\mu _4 -3$) depend on the first and second power of the ratio of second derivative of $\bar T$ and the first derivative of $\bar T$ (embedded in $\sigma _T$), respectively. Therefore, response of skewness and kurtosis to climate change depends on how the second derivative of the equilibrium temperature changes in relation to the first derivative. As in Eq. (\ref{moments_expanded}), this ratio is present in moments of any order, so the same discussion applies to any moment. 

Eq.(\ref{sigmaT_theory}-\ref{kurtosis}) are in agreement with past work by other authors. For example, Eq. (\ref{sigmaT_theory}) agrees with the expression of variance in \citet{schneider2015physics}. 
Lastly, Eq. (\ref{kurtosis}) is consistent with the expression of kurtosis in \citet{tamarin2022simple}, notably for the relationship 
\beq
\mu_4 = a\mu_3^2+b. \label{S_squared}
\eeq
In the same paper, Tamarin-Brodsky computed the coefficient $a$ in Eq. (\ref{S_squared}) from reanalysis data and obtained 1.6 in mid-latitude both in NH and SH. This is in agreement with our value of $a = 5/3 \approx 1.67$, calculated from Eqs. (\ref{kurtosis}) and (\ref{muT_theory}).



As indicated in Eq. (\ref{T-T2}), the time-mean temperature and the source temperature are not the same if the second derivative (rather, at least one derivative of an even order) of the equilibrium temperature profile is nonzero. 
However, since the lowest-order correction to the difference between time-mean temperature and equilibrium temperature is with the second derivative of the temperature, the expression of the standard deviation (\ref{sigmaT_theory}) happens to be unchanged from the well-accepted expression (\ref{eq:schneider}) suggested by \cite{schneider2015physics}. This difference does yield corrections to any moment that contains a derivative of order two or higher, including skewness and excess kurtosis, albeit solely in the numerical factor.

Additionally, Eq. (\ref{T-T2}) also suggests a physical interpretation of the source field. Specifically, it shows that the time-mean temperature can be interpreted as the solution to a diffusion equation, similar to that used in zonal-mean energy balance models (e.g., \cite{north1975theory}). This suggests that the source temperature can be interpreted as the radiative equilibrium temperature in an atmosphere with no horizontal mixing.

\section{Climate model analysis}

\begin{figure}
\centering
    \includegraphics[width=0.98\textwidth]{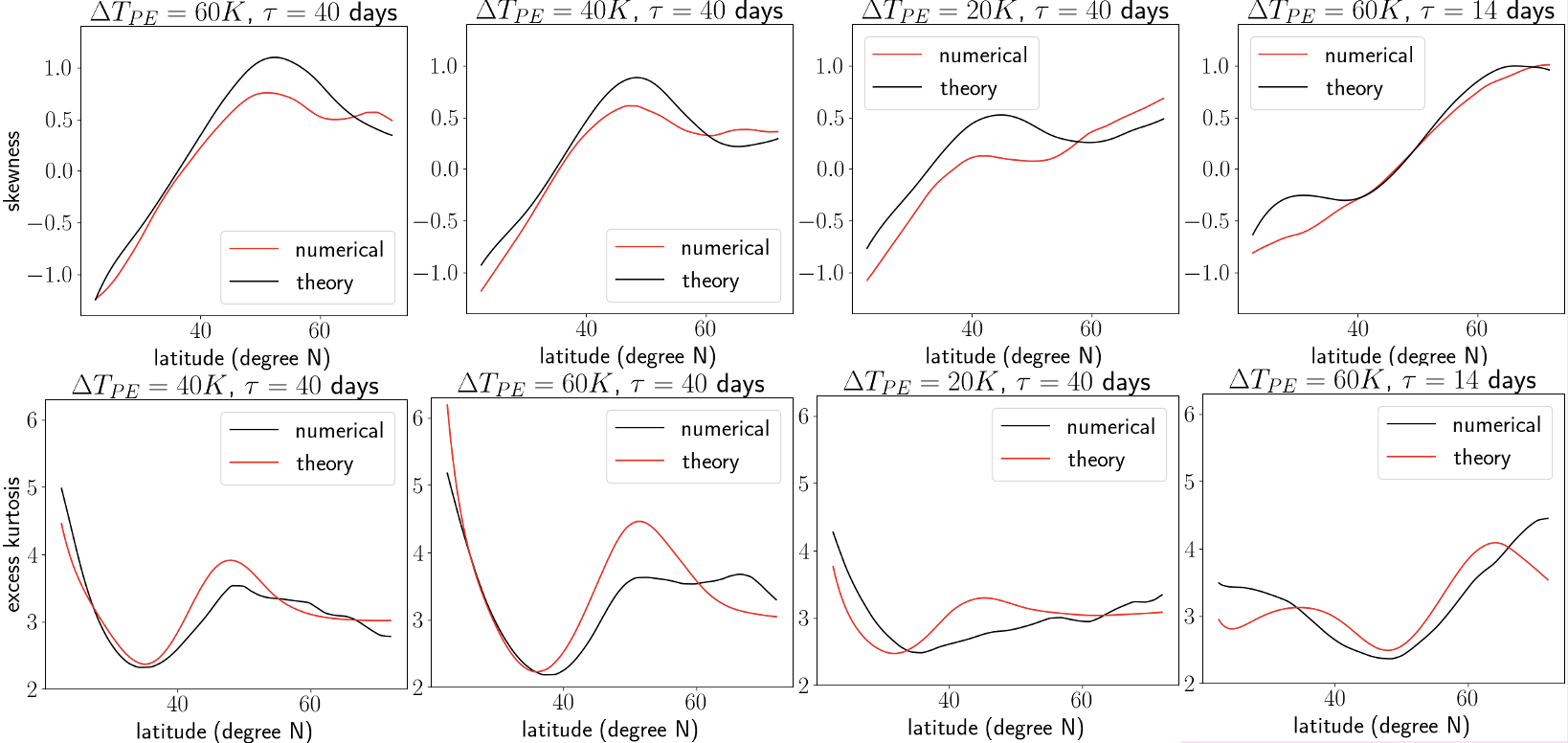}
    \caption{Skewness (top row) and excess kurtosis (bottom row) of temperature distribution in ISCA model for the cases where pole-to-equator temperature gradient $\Delta T_{PE}$ is 60$^\circ$K and relaxation time $\tau$ is 40 days (fitted value of $\sigma_\eta = 6.95$ degrees), $\Delta T_{PE} =$ 40$^\circ$K and $\tau = $ 40 days (fitted value of $\sigma_\eta = 6.22$ degrees), $\Delta T_{PE} =$ 20$^\circ$K and $\tau = $ 40 days (fitted value of $\sigma_\eta = 5.87$ degrees), and $\Delta T_{PE} =$ 60$^\circ$K and $\tau =$ 14 days (fitted value of $\sigma_\eta = 6.86$ degrees). Red lines show values calculated directly from the numerical model, and black lines show values calculated via Eq. (\ref{muT_theory}) or (\ref{kurtosis}) }
\label{ISCA_moments_figure}
\end{figure}

We test our theory using the ISCA \citep{gmd-11-843-2018} atmospheric General Circulation Model (GCM), in a dry-core \citet{held1994proposal} configuration at T85 resolution, run for 20 years. To compare our theory with the data, we calculate standard deviation, skewness, and kurtosis from data directly and via Eqs. (\ref{sigmaT_theory}-\ref{kurtosis}) as functions of latitude. The value of $\sigma_\eta$ is computed by fitting standard deviation of temperature calculated by Eq. (\ref{sigmaT_theory}) with the same quantity calculated from ISCA data via 20 points (about 28 degrees) that are closest to the location that has the maximum value of $\sigma_T$ in ISCA data. In all chosen cases, maximum value of $\sigma_T$ was located between 35 degrees and 45 degrees. A single calculated value of $\sigma_\eta$ is used in Eq. (\ref{muT_theory}) and Eq. (\ref{kurtosis}) to predict the values of skewness and kurtosis at all latitudes. 

In order to understand how temperature distributions might change in a changing climate, we evaluate our theory against idealized ISCA GCM runs with different equator-to-pole temperature gradients $\Delta T_{PE}$ and different RCE equilibration time scales $\tau$. Figure \ref{ISCA_moments_figure} shows the comparison of theoretical moments with those calculated directly from ISCA output. Predictions of moments based on the conceptual model show good agreements in mid-latitudes when pole-to-equator temperature $\Delta T_{PE}$ is comparable to that with the current Earth ($\sim 40^\circ$K), but theory and data deviate when $\Delta T_{PE}$ is as small as 1/2 of the current value (Fig. \ref{ISCA_moments_figure}). This indicates that nonliniarity in equilibrium temperature profile plays a major role in the shape of atmospheric temperature distribution in mid-latitudes with the current state of the atmosphere on Earth, but the dominant contribution to the atmospheric distribution most likely switches to a different physical process as $\Delta T_{PE}$ decreases, such as to a change in meridional displacement. This further suggests that our theory is  more accurate in winter than in summer, as  $\Delta T_{PE}$ is larger in winter. On the other hand, even though changing the value of the relaxation time $\tau$ has a large impact on the shape of skewness and kurtosis, it does not affect the consistency between theory and GCM in the mid-latitudes.

\begin{figure}
\centering
\includegraphics[height = 9 cm]{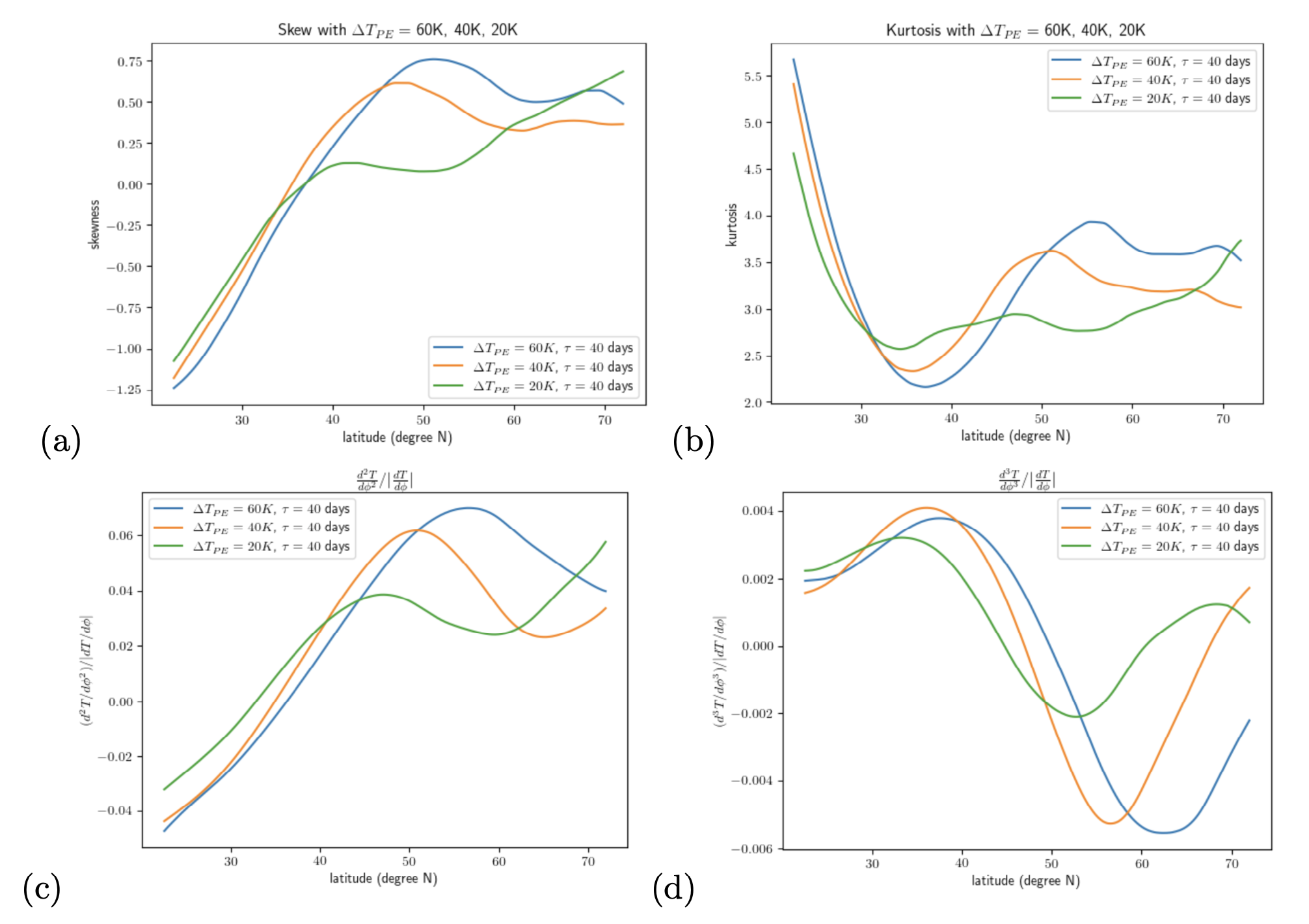}
\caption{(a) Skewness, (b) kurtosis, (c) the ratio $(d^2 \bar T/d\phi^2)/(d \bar T/d\phi)$, and (d) the ratio $(d^3 \bar T/d\phi^3)/(d \bar T/d\phi)$ in ISCA data with various pole-to-equator temperature gradients. Blue lines are for $\Delta T_{PE} = 60^\circ$K, orange lines are or $\Delta T_{PE} = 40^\circ$K, and green lines are for $\Delta T_{PE} = 20^\circ$K.}
\label{skew_kurtosis_deltaT}
\end{figure}

Our theory also allows us to calculate how much of the change in skewness and kurtosis comes from changes in temperature profile $\bar T (\phi)$, and how much comes from the changes in the displacement length $\sigma_\eta$. As in Eqs. (\ref{muT_theory}) and (\ref{kurtosis}), the change in moments due to change in external conditions such as $\Delta T_{PE}$ can come from the change in standard deviation $\sigma_\eta$ of the wind distribution and from the change in temperature profile--ratios of temperature derivatives ($(d^2 \bar T/d\phi^2)/(d \bar T/d\phi)$ for skewness, and both $(d^2 \bar T/d\phi^2)/(d \bar T/d\phi)$ and $(d^3 \bar T/d\phi^3)/(d \bar T/d\phi)$ for kurtosis). Our analysis shows that values of $\sigma_\eta$ for $\Delta T_{PE}$ = 60 K, 40 K, and 20 K are 6.95 degrees, 6.22 degrees, and 5.87 degrees, respectively. Changes in $(d^2 \bar T/d\phi^2)/(d \bar T/d\phi)$ and $(d^3 \bar T/d\phi^3)/(d \bar T/d\phi)$ as well as changes in skewness and kurtosis with three different values of $\Delta T_{PE}$ are shown in Figure \ref{skew_kurtosis_deltaT}. We find that the effects of the two sources are comparable with change in pole-to-equator temperature $\Delta T_{PE}$: when $\Delta T_{PE}$ changes from $60^\circ$K to $40^\circ$K, the maximum value of skewness changes by about 20 \%, where about 10 \% of that change comes from the change in $\sigma_\eta$, and another 10 \% is from change in the derivative ratios. On the other hand, the change in relaxation time changes the shape of the derivatives greatly without changing the value of $\sigma_\eta$ significantly. When $\tau$ changes from 40 days to 14 days, $\sigma_\eta$ changes from 6.95 degrees to 6.86 degrees, but the shape of skewness and kurtosis changes significantly as seen in Figure \ref{ISCA_moments_figure}.

\section{Summary and conclusions}

In this work, we derived analytical expressions of n\textsuperscript{th}-order temperature moments that arise from nonlinearity of the time-mean temperature profile in latitude. The expression for the moments is a function of derivatives of time-mean temperature and mean meridional displacement of air. We further derived simple closed-form expressions for variance, skewness, and kurtosis  to the lowest order in displacement of air $\eta$. In our expressions, variance is determined by the first derivative of background temperature, which is consistent with past work by \cite{schneider2015physics}. Skewness is determined by the second derivative of background temperature. This means that positive curvature of temperature leads to positive skewness, as is the case near poles. Similarly, negative curvature of temperature leads to a negative skewness in tropics and subtropics. Kurtosis is determined by the first three derivatives, which are consistent with \cite{tamarin2020changes}, (2022), including the coefficient $a$ in the expression that relates skewness $S$ and kurtosis $K$, $K = aS+b$ for mid-latitudes.


Since our method assumes that temperature variability comes only from advection, the methods are likely to be applicable to wintertime temperature given the minimal influence of surface fluxes to atmospheric temperature. Our theory deviates from the ISCA data as pole-to-equator temperature gradient gets greatly reduced. Thus, it is likely that the dominant contribution to the non-Gaussian distribution function changes to another process with climate change, and more research is needed to include such effects. 

Our theory is able to predict the moments of temperature distribution in the mid-latitudes in a Held-Suarez atmospheric model under a variety of equator-to-pole temperature gradients. The Held-Suarez  setup contains many of the physical mechanisms that have been suggested as potentially contributing to non-Gaussianity, such as geometric effects, nonlinear advection, and asymmetry in meridional displacement of air parcels \citep{garfinkel2017non}. The fact that our theory is able to reproduce the higher order moments and their changes in a changing climate suggests suggests that  nonlinearity in background temperature gradient is the dominant contribution for non-Gaussianity of observed atmospheric temperature distributions in the mid-latitudes. 

The Held-Suarez setup also contains atmospheric dynamics such as jet stream and Rossby waves that have been argued to be important in changing temperature distributions. In particular, it has been suggested that a decreased equator-to-pole temperature gradient would lead to a weakening of the jet stream, which in turn would yield larger displacements of air and increase the frequency and/or intensity of cold-air outbreaks in winter \citep{francis2012evidence}. Contrary to this hypothesis our results show that the effective mean absolute displacement $\sigma_\eta$ decreases with a decreasing equator-to-pole temperature gradient. 

Due to the use of large-scale physics to compute statistical moments of temperature, our work contributes to the understanding of how large-scale dynamics of atmosphere can affect local-scale temperature variability and extremes. 
Higher order moments are difficult to infer from the short observational record, especially in the presence of non-stationarity, such as arises from a changing climate. The difficulty is amplified if one is interested in \textit{changes} in these higher order moments and the associated changes in temperature extremes. Our theory offers a path for predicting changes in temperature extremes based on quantities for which there are robust observations and predictions, such as advection length scale and the time-mean meridional temperature profile.

\newpage

\section{Data Availability Statement}

The code for the ISCA climate model is publicly available at \\ \noindent https://execlim.github.io/IscaWebsite/index.html. The model was run in the standard Held-Suarez configuration for 20 years. 
The code to analyze ISCA model output and reproduce the figures is found at https://github.com/keikoino0212/Analytical-Model-for-the-Higher-Order-Moments-of-Midlatitude-Atmospheric-Temperature-Distributions.

\acknowledgments

This work was supported by the U.S. Department of Energy, Office of Science, Biological and Environmental Research Program, Earth and Environmental Systems Modeling, MultiSector Dynamics under Cooperative Agreement DE-SC0022141.  Any opinions, findings, and conclusions or recommendations expressed in this material are those of the authors and do not necessarily reflect the views of the funding agencies.

We thank Neutrina Kircher for her unwavering support.

\bibliography{citation.bib}

\end{document}